# Direct visualization and control of SrO$_x$ segregation on semiconducting Nb doped SrTiO$_3$ (100) surface

Hyang Keun Yoo,[1,2,3] Daniel Schwarz,[1] Søren Ulstrup,[1] Woojin Kim,[2,3] Chris Jozwiak,[1] Aaron Bostwick,[1] Tae Won Noh,[2,3] Eli Rotenberg[1,*] and Young Jun Chang[4,5,$]

[1]*Advanced Light Source (ALS), E. O. Lawrence Berkeley National Laboratory, Berkeley, California 94720, USA*

[2]*Center for Correlated Electron Systems, Institute for Basic Science (IBS), Seoul 08826, Korea*

[3]*Department of Physics and Astronomy, Seoul National University, Seoul 08826, Korea*

[4]*Department of Physics, University of Seoul, Seoul 02504, Korea*

[5]*Department of Smart Cities, University of Seoul, Seoul 02504, Korea*

*erotenberg@lbl.gov & $yjchang@uos.ac.kr

**ABSTRACT**

We investigated how $SrO_x$ segregates on a Nb doped $SrTiO_3$ (100) surface by in air annealing. Using atomic force and photoemission electron microscopes, we can directly visualize the morphology and the electronic phase changes with $SrO_x$ segregation. $SrO_x$ islands less than 2 $\mu$m in size and 1-5 unit cells thick nucleate first and grow in a labyrinth domain pattern. After prolonged annealing, $SrO_x$ forms a ~10 nm thick film. We show that the domain pattern can be controlled by introducing a surface miscut angle of $SrTiO_3$. Additionally, the segregated $SrO_x$ has a lower work function, compared to that of $SrTiO_3$. These results suggest that the control and tunability of $SrO_x$ segregation is applicable to the design of a new functional electronic devices in the semiconducting $SrTiO_3$ based heterostructure.

## I. INTRODUCTION

SrTiO$_3$ (STO) is one of the most popular substrate materials for transition metal oxide heterostructures[1-9]. This is not only due to the commercial supply of very high quality crystals, but also many interesting phenomena in STO based heterostructures[4-10]. For example, high-$T_c$ superconductivity has recently been discovered in a monolayer FeSe film on electron doped STO substrate[4]. Two-dimensional conducting electrons are formed at the interface between two band insulators, LaAlO$_3$ film and STO substrate[5,9]. These interesting phenomena are strongly dependent on the surface treatment method of the STO substrate[4,5]. An improved understanding and mehtods of manipulating the STO surface are of central importance. The recipes, which commonly consist in a combination of chemical acid etching and thermal annealing, to achieve the atomical A-site (SrO)[11-14] and B-site (TiO$_2$)[15-19] terminated surfaces have been intensively developed over 20 years. As a result, a TiO$_2$-termination induced by etching is relatively well-understood and routinely used nowadays, but the recipe to control a SrO-termination still remains elusive.

There are two well-known methods to achieve a SrO-covered STO surface: one is the deposition of a SrO monolayer on a TiO$_2$-terminated STO substrate[11,12] and the other is annealing of electron doped STO in air or with molecular oxygen gas[13,14,20]. In this letter, we elucidate how SrO$_x$ segregates on the surface of Nb doped STO (Nb:STO) crystals during air annealing. The morphology and the electronic phase changes with the segregation are directly visualized by using photoemission electron microscopy (PEEM) and atomic force microscopy (AFM). We demonstrate the total process, from the atomical SrO$_x$ segregation to the ~10 nm thick film formation, on the surface by annealing, as well as the controllability of the segregation by varying the surface miscut angle and electron doping concentration. We expect that our results will provide novel insights, not only to develop the recipe for the SrO-terminated STO surface, but also to design a new STO based heterointerface with $SrO_x$ segregation control.

## II. EXPERIMENT

We prepared several different Nb:STO substrates by varying the surface treatment recipes. Before annealing atomically flat substrates were prepared by buffered-hydrofluoric acid (BHF) etching to create $TiO_2$-terminated Nb:STO surfaces[15-17]. Then, we annealed the substrates from 1000 to 1300°C for 2 to 72 hours. Additionally, we compared the surface of Nb:STO with different surface miscut angles and Nb doping concentration using the same preparation. Note that the BHF etching was applied to all samples and the annealing was performed in air, except when *in vacuum* annealing is explicitly mentioned.

The PEEM, work function ($W_F$) and x-ray photoemission spectroscopy (XPS) measurements were performed at the MAESTRO facility at Beamline 7.0.1 of the Advanced Light Source[21]. The PEEM images were obtained by exciting photoelectrons using the ultraviolet (UV) light of a Hg arc discharge lamp (~4.5 eV). The brightness in the PEEM images is dominantly determined by the intensity of the secondary electrons. The $W_F$ of Nb:STO is around 3.9 eV[22]. This implies that a different brightness in image stems from a local difference in $W_F$ of the sample. The quantitative local $W_F$ value was measured in the PEEM instrument using the secondary electron cut-off with a He-I (21.2eV) source. The XPS was measured in the micro-ARPES endstation in total-energy resolution (photons + electrons) of around 50 meV, corresponding to photon energy of 240 eV at around 90 K. The measurement power density of the photon source was kept at 0.5 W cm$^{-2}$, relatively week intensity. We found that the electronic structure of sample surface was kept stable under the UV irradiation. Note that the Nb:STO shows a significant UV irradiation effect, such that much intense UV power density of ~ 4 W cm$^{-2}$ could generate oxygen vacancies on the STO surface[8,23-26], but the cation stoichiometry is not considerably affected. The non-contact mode AFM was performed in air and at room temperature.

III. RESULTS AND DISCUSSION

We found that the island structures having a different $W_F$ value compared to that of Nb:STO appear on surface after annealing. Figures 1(a) and 1(b) show PEEM images of 0.5 wt% Nb:STO substrates after etching and 72 hours of annealing at 1000°C (0.5Nb-1000C-72h), respectively. While the just etched sample is homogeneous, we observe dark spots with a different $W_F$ appear after annealing (cf. Fig. 1(b)). To get structural information in the dark spots, we examined both samples with AFM. The just etched sample has a very flat surface in Fig. 1(c), while the 0.5Nb-1000C-72h sample shows a clear step structure in Fig. 1(d). We obtained the line profile in Fig. 1(e) along the red dashed line shown in Fig. 1(d). The line includes a step induced by annealing. The step heights are around 200 and 600 pm which are not matched with the 400 pm unit cell of Nb:STO[15,16] or its multiples.

The dark spots can be attributed to several unit cells of SrO$_x$ segregated by annealing. The A-site segregation easily occurs in the perovskite structure due to the favored Schottky-type disorder[20,26]. Additionally, the A-site segregation shows steps of half a unit cell or one and a half unit cells of the original perovskite structure on the surface[27,28]. The unit cell of SrO$_x$ is around 200 pm and its electronic structure is different from that of STO[27,28]. Thus, we can explain the two electronic phases observed in PEEM and the different step heights in the AFM images with the SrO$_x$ segregation on the surface. Additionally, the XPS results also support our picture of SrO$_x$ segregation on the Nb:STO surface after annealing. The Sr 3d spectra of just etched and 0.5Nb-1000C-72h samples in Figs. 1 (f) and 1(g) are fitted by two components: the bulk (blue) and the surface (green) components at around 133.8 and 134.5 eV binding energy, respectively. The surface component can be attributed to segregated Sr atoms on the surface[14,29,30], and it is stronger for the annealed sample.

In more annealing step, the segregated $SrO_x$ islands form a labyrinth domain pattern, and grow in a ~10 nm thick film by filling the gaps between domains. The PEEM image of 0.5Nb-1300C-2h in Fig. 2(a) shows the $SrO_x$ labyrinth pattern clearly. Then, as shown in 0.5Nb-1300C-12h in Fig. 2(b), the gaps between domains are filled and form a closed $SrO_x$ film with new islands nucleated on it. The AFM images in Figs. 2(c)-(e) also show a consistent $SrO_x$ segregation process. Additionally, in the line profiles along the red dashed lines in the bottom panels, we measure the heights of $SrO_x$ domains and films to be ~10 nm. The schematic diagrams of the segregation process, based on our experimental results, are illustrated in Fig. 3. This mesoscopic length scale pattern can be understood by elastic theory in the presence of interactions between domain boundaries[31-33]. A commonly accepted explanation of 2D islands on solid surfaces is based on the delicate balance between attractive forces between neighboring atoms and the long-range repulsive interactions due to surface stress. If two phases of different surface stress coexist, then the surface lower the surface free energy by forming striped domains. Therefore, the stripe width and periodicity are decided by the balance between the energy of forming linear domain boundaries between the two phases and the energy of the long-range substrate and surface elastic relaxations.[31] The substrate elastic anisotropy also influences the shape anisotropy of 2D islands.[32,33]

It is known that the change of domain pattern from an island to an elongated shape occurs with the elastic anisotropy stress on the surface[32,33]. We demonstrate the stripe domain pattern of the segregated SrO$_x$ by applying the surface miscut of Nb:STO. The Nb:STO substrates in Fig. 4 have a surface miscut angle around 0.1°, whereas the samples in Figs. 1, 2 has a miscut smaller than 0.05°. The PEEM and AFM images of 0.05Nb-1300C-12h [Figs. 4(a),(b)] exhibit a stripe domain pattern of SrO$_x$ and it becomes completely separated in two stripe domains in 0.05Nb-1300C-72h [Figs. 4(c),(d)]. We attribute the observed stripe domain pattern to the anisotropic surface stress due to surface miscut angle.

The low concentration of Nb doping induces less SrO$_x$ segregation due to low electron concentration in the STO[20]. The results of 0.05Nb-1300C-2h in Fig. 4 show less segregation, compared with that of 0.5Nb-1300-2h in Fig. 2. There is the surface miscut angle difference between both samples, but we also confirmed that 0.05Nb-1300C-12h, having a surface miscut angle smaller than 0.05°, did not exhibit the big SrO$_x$ islands shown in Fig. 2 (b) (not shown). These results suggest the Nb doping concentration proportional to the electron concentration is more important to determine the amount of the SrO$_x$ segregation.

We investigated *in vacuum* annealing effect for the segregated SrO$_x$. The segregated islands in the pristine 0.5Nb-1100C-2h sample show darker color than the surrounding Nb:STO areas in PEEM [Fig. 5(a)]. However, after *in vacuum* ($10^{-9}$ Torr) annealing at 150° C for 10 minutes, the contrast between the segregated SrO$_x$ islands and the other Nb:STO areas reverses [Fig. 5(b)]. The SrO$_x$ areas becomes brighter than that of Nb:STO. The SrO$_x$ is not a stable phase in air, so many adsorbates can be attached to it[34,35] and detach by *in vacuum* annealing. To confirm our hypothesis, we measured the core levels before and after *in vacuum* annealing. The O 1s core level in Fig. 5(c) shows a clear change. The spectra in the higher binding energy range are attributed to the surface adsorbates[36]. The *in vacuum* annealing reduces the spectral weight in the higher binding energy range, which indicates the removal of the surface adsorbates. This can explain the change of contrast in Figs. 5(a) and 5(b). Note that we confirmed that more adsorbates are attached to the segregated $SrO_x$ in air than that of Nb:STO by comparing with the O 1s core level of just etched sample (not shown).

We also measured the quantitative $W_F$ value for the clean $SrO_x$. We sequentially measure the local $W_F$ values as going farther from the $SrO_x$ island. The measurement positions are marked in Fig. 5(b) by squares. Then, we deposited a cobalt (Co) metal on the Nb:STO area to get a reference for the secondary electron cut-off energy value. The spectral line of Co/Nb:STO is the reference shown by black dashed line in Fig. 5(b). The experimental result exhibit that the $W_F$ of clean $SrO_x$ is ~0.5 eV lower than that of Nb:STO[37], so the $SrO_x$ segregated area has a different heterojunction alignment with Co metal compared to that of Nb:STO. This result indicates that the segregated $SrO_x$ can be applicable for new component to design a functional electronic device in semiconducting STO based heterostructure.

## IV.  CONCLUSION

In summary, we investigated how $SrO_x$ segregates on the surface of Nb:STO with annealing by using PEEM, XPS and AFM techniques. Our results provide the total process of $SrO_x$ segregation from atomic level to ~10 nm thick film formation. We also demonstrated the controllability of the segregation by tuning the surface miscut angle and the electron doping level of STO. The $W_F$ of $SrO_x$ are also different from that of Nb:STO. A clear implication of our work is that the electronic properties at the semiconducting STO based heterointerfaces can be tuned by control of the $SrO_x$ segregation.

**ACKNOWLEDGEMENTS**

The Advanced Light Source is supported by the Director, Office of Science, Office of Basic

Energy Sciences, of the U.S. Department of Energy under Contract No. DE-AC02-05CH11231. This work was supported by IBS-R009-D1. D.S. acknowledges financial support from the Netherlands Organisation for Scientific Research under the Rubicon Program (Grant 680-50-1305). S.U. acknowledges financial support from the Danish Council for Independent Research, Natural Sciences under the Sapere Aude program (Grant No. 4090-00125). Y.J.C. acknowledges support from National Research Foundation of Korea (NRF-2019K1A3A7A09033389, 2020R1A2C200373211) and [Innovative Talent Education Program for Smart City] by MOLIT.

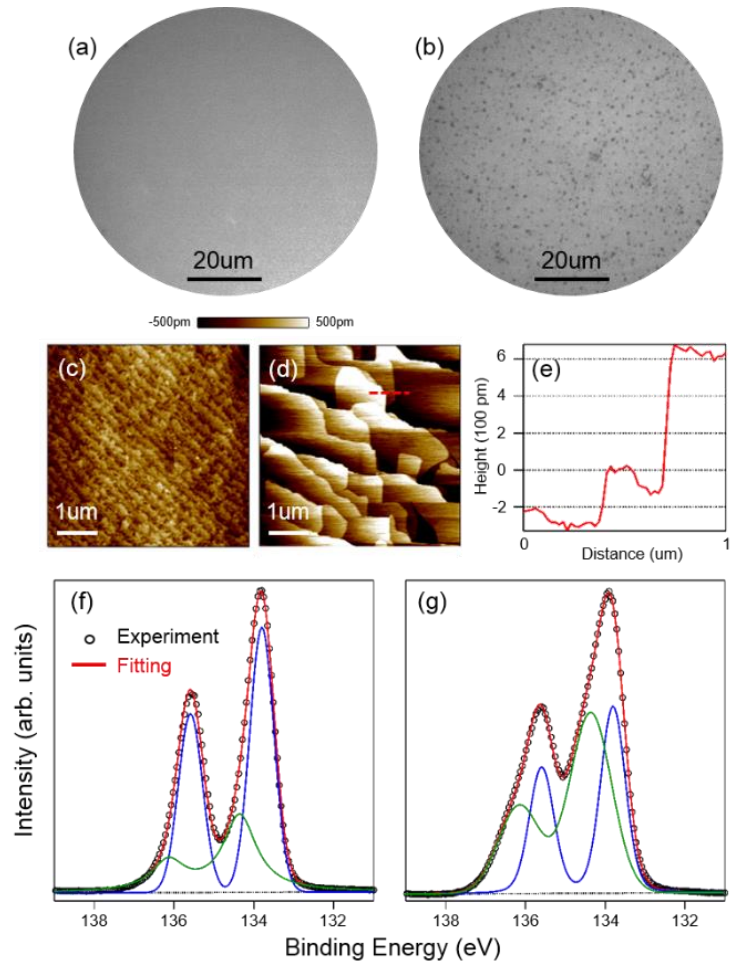

FIG. 1. Photoemission electron microscopy (PEEM) images of 0.5 wt% Nb doped SrTiO$_3$ (Nb:STO) after (a) buffered-HF (BHF) etching and (b) BHF etching followed by 72 hours of annealing at 1000°C. (0.5Nb-1000C-72h). Atomic force microscopy (AFM) images of (c) etched and (d) 0.5Nb-1000C-72h samples. (e) Line profile along red dashed line in (d). Sr 3d core level spectroscopy results for (f) etched and (g) 0.5Nb-1000C-72h samples. The fits of the bulk (blue) and the surface (green) components are included.

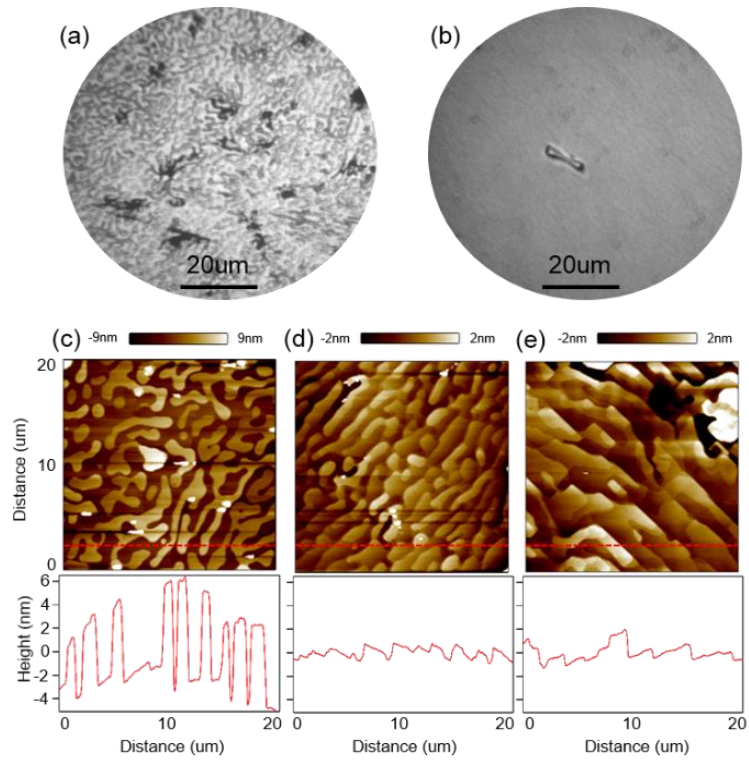

FIG. 2. PEEM images of (a) 0.5Nb-1300C-2h and (b) 0.5Nb-1300C-12h samples. AFM images of (c) 0.5Nb-1300C-2h, (d) 0.5Nb-1300C-6h and (e) 0.5Nb-1300C-12h. The line profiles along the red dashed lines in (c)-(e) are shown in the bottom panels. Note that the AFM in 0.5Nb-1300C-12h sample is measured in the area far from the island shown in (b).

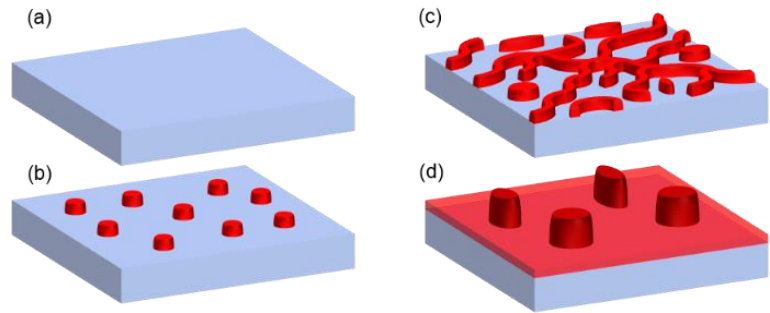

FIG. 3. Schematic diagrams of SrO$_x$ segregation on Nb:STO. (a) TiO$_2$ terminated Nb:STO after BHF etching. (b) SrO$_x$ islands nucleation and (c) labyrinth pattern of its domain in annealing. (d) SrO$_x$ film formation, and appearance of new big islands on the film after more annealing steps.

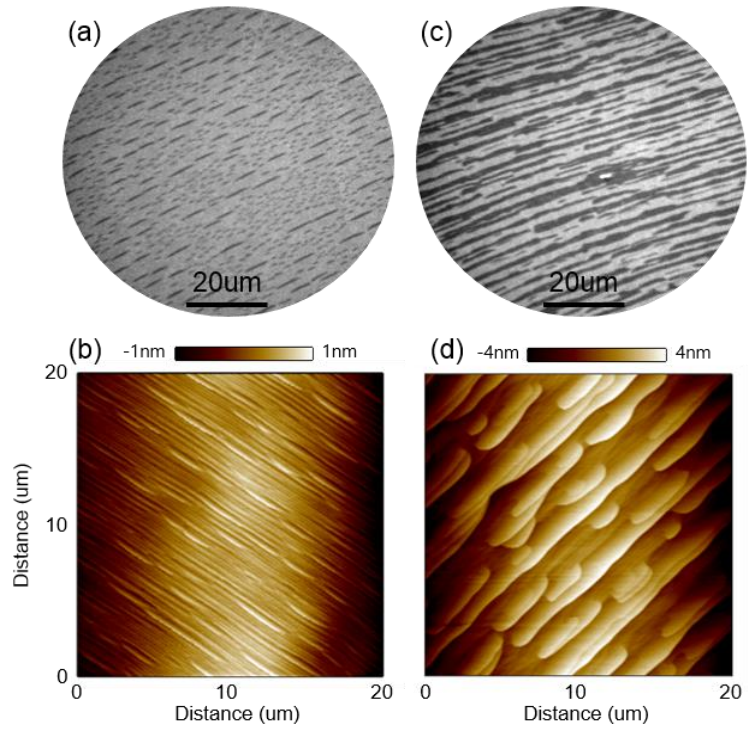

FIG. 4. PEEM and AFM images of (a),(b) 0.05Nb-1300C-12h and (c),(d) 0.05Nb-1300C-72h samples. The SrO$_x$ domain has a stripe pattern. This sample has a surface miscut angle ~0.1°, while the samples in Figs. 1, 2 have it < 0.05°.

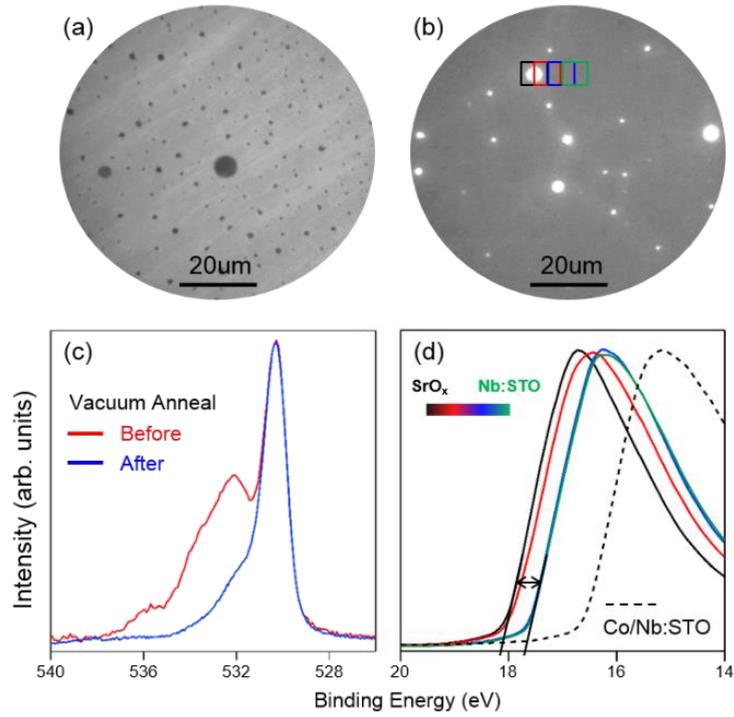

FIG. 5. PEEM images of 0.5Nb-1100C-2h in (a) pristine condition and (b) after additional *in vacuum* annealing at 150°C for 10 minutes. (c) O 1s core level before and after *in vacuum* annealing. (d) Measurement of work function ($W_F$) for SrO$_x$. The measurement positions are marked in (b) by squares. The $W_F$ of clean SrO$_x$ is ~0.5eV lower than that of STO, as shown by a double-headed arrow. A cobalt (Co) metal film on Nb:STO is shown as a reference.

18